\documentclass[sn-aps]{sn-jnl}
\usepackage{graphicx}%
\usepackage{multirow}%
\usepackage{amsmath,amssymb,amsfonts}%
\usepackage{amsthm}%
\usepackage{mathrsfs}%
\usepackage[title]{appendix}%
\usepackage{xcolor}%
\usepackage{textcomp}%
\usepackage{manyfoot}%
\usepackage{booktabs}%
\usepackage{algorithm}%
\usepackage{algorithmicx}%
\usepackage{algpseudocode}%
\usepackage{listings}%

\begin{document}

\title[Detection of presence and number of persons by a Wi-Fi signal: a practical RSSI-based approach]{Detection of presence and number of persons by a Wi-Fi signal: a practical RSSI-based approach}
\author*[1]{Dario Juki\'{c}}\email{djukic@grad.unizg.hr}
\author[2]{Silvije Domazet}
\author[3]{Ante Ivanko}
\author[3]{David Raca}
\author[3]{Sini\v{s}a Nikoli\'{c}}
\author[3]{Marin Kne\v{z}evi\'{c}}
\author[3]{Filip Jovi\'{c}}
\author[3]{Nenad Raca}
\author[2]{Hrvoje Buljan}
\affil[1]{Faculty of Civil Engineering, University of Zagreb, A. Kačića Miošića 26, 10000 Zagreb, Croatia}
\affil[2]{Department of Physics, Faculty of Science, University of Zagreb, Bijenička c. 32, 10000 Zagreb, Croatia}
\affil[3]{Aduro Idea, Zavrtnica 17, 10000 Zagreb, Croatia.}

\abstract{
We present experimental results and theoretical methods for the precise determination of the presence and the number of persons in an observed area by using Wi-Fi signals. Our setup does not require active cooperation of persons present in the Wi-Fi field, and relies only on the Received Signal Strength Indicator (RSSI), which is read by the detectors. We first show that the standard deviation of the measured RSSI data can be used as a practical tool to establish the presence of a person (or more persons) with high precision, in particular when the signal source is inside the measurement room. For the more difficult problem of counting the number of persons, we have employed machine learning algorithms to analyze data collected on nine different detectors and up to nine people present in our experiment. We have achieved excellent results (prediction accuracy of $98 \%$ and above) for counting already with only few detectors utilized in the analysis.
While generalizations to nontrivial indoor geometries (such as odd shapes, more rooms, and greater size) may be of interest in some applications, where additional care with respect to positioning od detectors can be needed (or even placing additional detectors), this approach may be useful due to its conceptual practicality and solid prediction results.
}


\maketitle

\section{Introduction}
\label{sec:introduction}

The relevance and omnipresence of wireless networking in everyday life have also led to important developments in sensing human activities, with Wi-Fi signals generally providing cost-effective and non-invasive alternative to other sensor technologies.
As the rich plethora of applications constantly expands, the Wi-Fi sensing today may include human (of object) presence detection, localization, tracking, counting of people, monitoring their vital signs, recognition of daily activities, gesture recognition, and human identification and authentication (for a comprehensive review, see, e.g., Refs. \cite{Ma2019,Liu2020}).
Broadly speaking, the Wi-Fi sensing relies on measuring changes of the Wi-Fi signal due to the presence or activity of one (or more) persons in an area chosen for monitoring.
In order to quantify these changes, several approaches have been utilized.
The Received Signal Strength Indicator (RSSI) measured by the Wi-Fi detector may be considered as a basic quantity of interest as it provides information only on the intensity of the electromagnetic radiation. Yet, being easily available it has been employed since early studies in the Wi-Fi sensing field \cite{Youssef2007,Moussa2009,Kosba2012,Yuan2011,Xu2013}.
Channel State Information (CSI) gives more information on the signal, as is contains data on both amplitude and the phase for each of multiple frequencies (subcarriers) in a given bandwidth.
Because of this, it presents more advanced and reliable approach which is being employed essentially in all domains of Wi-Fi sensing \cite{Ma2019,Liu2020}.
Parallel to this, more specialized (and customized) techniques have been developed which are based on the frequency shift of radio waves reflected by human body. More specifically, the Frequency Modulated Carrier Wave (FMCW) method, which allows for measuring the depth of a reflector, has been used in studies which include 3D body tracking \cite{Adib2014}, capturing human figure through a wall \cite{Adib2015m1}, monitoring vital signs \cite{Adib2015m2}, recognizing emotions \cite{Zhao2016}, and monitoring sleep and insomnia \cite{Hsu2017}.
Futhermore, the Doppler shift of radio waves has been utilized for sensing through the wall \cite{Chetty2012, Adib2013} in order to detect human motion, for gesture recognition \cite{Pu2013,Tan2014}, or for vital signs detection \cite{Li2016} in passive Wi-Fi radar (PWR) systems.
A recent study provides detailed comparison between the PWR and the CSI approach \cite{Li2022}.

\subsection{Presence detection and counting persons: RSSI and CSI studies}

The usage of Wi-Fi signals for establishing the presence of one or more persons inside a room or an open space can be of great practical interest, in particular when no active cooperation of persons is assumed (people do not have to carry their own Wi-Fi devices). A more challenging problem is that of counting the exact number of such persons, and to date it has been studied in dominantly two different directions. The first approach for counting uses only information about the signal strength on the detector, i.e,. the RSSI value read by the detector. The CSI provides for an alternative route, as it offers a much richer description of the signal, but is at the same time much more complex in practical terms.
We now first provide with a brief overview of relevant works in both directions (with emphasis on RSSI-based studies).

The first studies which introduced the concept of device-free passive detection and which were based only on the RSSI value of the detectors have utilized moving average and moving variance techniques \cite{Youssef2007}, or maximum likelihood estimator algorithm \cite{Moussa2009}. A robust RSS system for human motion detection has been presented in Ref. \cite{Kosba2012}. The system employed a non-parametric statistical anomaly detection technique which was shown to outperform former approaches.
Furthermore, inital studies have also investigated the level of presence (density) of people in an indoor environment \cite{Yuan2011}, or determined the exact number and location of small number of people (up to four people in Ref. \cite{Xu2013}).

Additional progress in determining the exact number of people using RSSI signals was reported by Depatla \emph{et al.} \cite{Depatla2015} (for patent application, see \cite{Depatla2017}). Only one pair of antennas (transmitter and detector) was placed stationary in the space, and there were people walking between them. Mathematical analysis of two important effects by which people influence the received signal strength was made. The first is blocking the direct line of sight (LOS) when the individual is exactly on the transmitter-detector line, and the second refers to the effects of signal scattering on people. The result of this analysis is an expression for the probability distribution of the received signal as a function of the number of persons present. By comparing the theoretical model and experimental results, the authors calculated the probability of success in counting people. For antennas (radiating in all directions) in closed spaces, the accuracy of counting up to a total of nine people was $55\%$ (with an error of up to one person), or $63\%$ (with an error of up to two people). In the outer space, these accuracies amounted to $64\%$ and $96\%$, respectively. Even better results were achieved by using directional antennas. In their subsequent works, they have also demonstrated the possibility of using RSSI data to determine the average walking speed of people in space \cite{Depatla2018a}, and for counting people through walls  \cite{Depatla2018b}.

We highlight also the work \cite{Yoshida2015} in which the authors collected the results of RSSI measurements at ten measuring points (detectors) inside one room, and the number of people was determined by using regression-based methods. In the first case they used the linear, and in the second the so-called support vector regression (SVR). SVR is a nonlinear regression method based on the Support Vector Machine algorithm, which is extremely popular in machine learning for solving classification and regression problems. The authors used these methods, among other things, to predict the presence of people in the room (people are present or not), as well as to determine the exact number of people. The SVR method proved to be more successful: in experiments with a maximum of seven people in the room, the accuracy of predicting presence was $98\%$, and determining the correct number of people was $77\%$.


A fine-grained approach to determine the presence and the number of people in an environment uses CSI data recorded by modern Wi-Fi systems. In this case, the Multiple Input - Multiple Output (MIMO) systems in Wi-Fi technology enable the information about the amplitude and phase of the signal to be recorded on many channels (frequencies). The channels are orthogonal to each other thanks to the Orthogonal Frequency Division Multiplexing (OFDM) technique, because the frequencies are chosen in such a way as to avoid interference during transmission (see, for example, Ref. \cite{Goldsmith}). 

By now, numerous works have identified the CSI as a more accurate and robust approach for presence detection \cite{Nishimori2011,Hong2012,Xiao2012,Xiao2013,Zhou2013,Honma2013,Qian2014,Wu2015}.
As for the problem of counting people, the first paper in which this approach has been used was published in 2014 \cite{Xi2014}. Based on experiments (in which up to 30 people participated) and the Gray Verhulst model, the authors determined the accuracy of predicting the number of people. In indoor spaces, the accuracy was about $80\%$ (with an error of up to one person), and $98\%$ (with an error of up to two people). In outdoor conditions, the model was less successful, with an accuracy of about $70\%$ (with an error of up to two people).
Subsequent papers have used different methods, mainly using machine learning techniques, to predict the number of people based on CSI data \cite{DiDomenico2016, Liu2017, Cheng2017, Liu2019, Mabuchi2020, Brena2021}. In addition to determining the number of people, the CSI offers many other potential applications, which are described in more detail in the review article~\cite{Ma2019}.

\subsection{Summary of our findings}

In this paper, we report on the measurements we have conducted and their analysis, in order to find practical (efficient) algorithms for detecting the presence and counting the number of people without their active cooperation, based only on the RSSI values read by detectors. We show that the standard deviation of the measured time series can be used as a feature which discriminates between the presence of one (or more persons) and the noise in an observed environment. Based on this conclusion, we define algorithms which have been trained only on the noise data and have shown excellent prediction accuracy for tested data samples, which is comparable to established (and more demanding) machine learning algorithms. However, we emphasize that the prediction quality can be somewhat reduced if the radiation source is located outside the measurement area. In the second part of the work, we have studied a more challenging problem of people counting, for this we have performed measurements with nine detectors and up to nine people in the measurement area. We have demonstrated that for this setup the well-known machine learning algorithms (e.g., Random Forest) can be used for classification purpose with excellent prediction results ($98 \%$ and above), already when data from only few (e.g., four) detectors are used.

\section{Methods/Experimental}

\subsection{Basic concepts of experiments and their theoretical analysis}

Our experiments and theoretical analysis have been conceived and carried out as follows. One room for experiments has been planned, the Wi-Fi signal source and Wi-Fi radiation detectors are located in that room. Alternatively, the source (and even the detectors) could also be placed outside the measurement room. Measurements were made by recording the RSSI values on the detectors.
For given positions of the source and detector, without the presence of people or any other moving objects, the radiation intensity measured on the detectors is constant with certain fluctuations (noise) which are a standard part of all measurements. Noise analysis will prove extremely important for the understanding and interpretation of measurement results.

When there are one or more people in the room, standing or moving, the radiation intensity on the detectors will change in relation to the noise value. Since the organs of the human body have a high dielectric permittivity of the order of $50-80$ compared to vacuum at Wi-Fi frequencies \cite{Gabriel1996}, people standing or moving scatter and absorb Wi-Fi radiation. These reflections and absorptions cause a detectable change in radiation intensity over time.
Based on the collected data, i.e., time series of radiation intensity, the key part of the work is to answer
(a) whether someone was in the area where the measurements were being taken, and
(b) how many people were present?
Various methods can be used for this, from statistical analysis of measured time series with and without people (and then building a predictive model), to machine learning methods where the algorithm learns from previously measured signals to make a prediction in real time. Furthermore, it may be useful to examine with what reliability different methods work, what kind of algorithm calibration is needed for a reliable result, and how the results change when the source or detectors are moved to other positions. Which positions of sources and detectors (and their number) are optimal for covering a certain space? The answers to these questions will be discussed in the following chapters.

\subsection{Experiments: detecting the presence of people}

\begin{figure}[!t]
\centerline{\includegraphics[width= 0.75 \columnwidth]{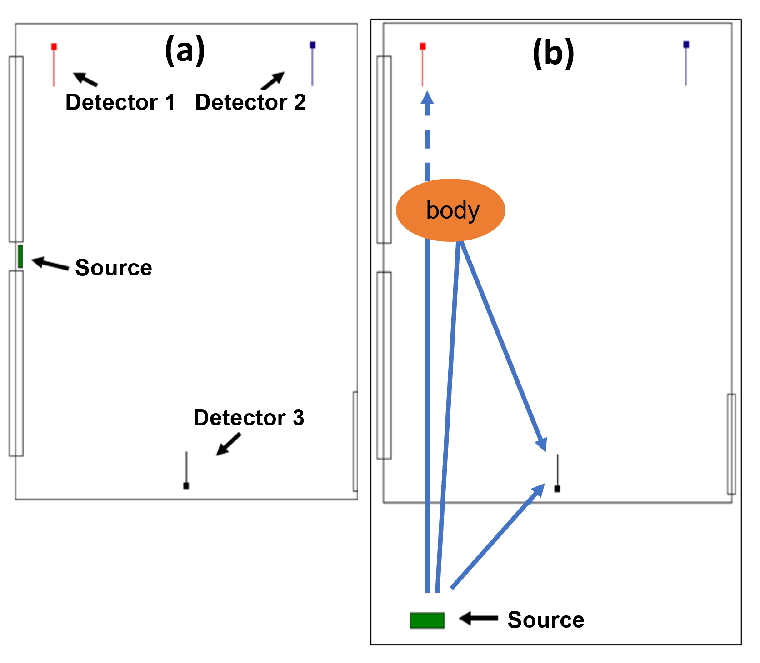}}
\caption{Floor plan of the measurement area and the position of the source and detectors during (a) the first and (b) the second set of measurements. The dimension of the measurement room is $3.5 \times 4.5$ m$^2$. (a) The source and detectors are inside the room. (b) The source is moved outside the room, while the positions of the detectors are not changed. Illustration of two processes that lead to changes in the intensity of the measured signal: (i) reflection and (ii) absorption by the human body. The signal in detector 1 is affected by radiation absorption, while the signal in detector 3 is affected by radiation reflection.}
\label{MjerniProstor}
\end{figure}

In this section we present two sets of experiments, both being setup with one source and three detectors.  
However, the spatial configuration of the source has been different: for the first set of measurements (labeled as M1) the source was inside the room where the measurements have been made, while for the the second set (labeled as M2) the source was placed outside.
Schematic illustration of the room and location of the source and the detectors for both sets of experiments are shown in Fig. \ref{MjerniProstor}.
We note that position of each detector relative to the source will affect whether a given detector measures more absorption or reflection of Wi-Fi radiation.
In principle, a more complex geometry of the indoor enviroment could be set, for example, an oddly shaped room, or multiple rooms covered by detectors. However, in that case a careful choice of detector positions should allow for similar analysis, with any particular detector being inside the same room or separated by a wall (or doors) from the source.

Let us consider the configuration of the source, detector and human body graphically shown in Fig. \ref{MjerniProstor}(b). Here, the human body is between the source and the detector 1 (D1); therefore, the measured radiation intensity at D1, relative to the noise, will show that the radiation has been absorbed. However, if we consider the positions of the source and the detector 3 (D3), deviation from noise radiation on D3 will be caused by radiation reflection.

Obviously, this is a simplified picture of geometric optics that does not take into account the wave nature of Wi-Fi radiation and possible interference from radiation reflected from other objects. Detectors measure all possible contributions and therefore, the analysis of measured signals also includes all possible contributions. For an accurate theoretical description of the system it is necessary to solve Maxwell's equations in a given space with adequate boundary conditions and an approximately accurate dielectric response of the human body to Wi-Fi radiation. It should be noted here that a simplified picture of absorption and reflection in geometric optics can still give a quick answer on how to place sources and detectors in a given space to maximize the desired precision of reading information from a Wi-Fi signal, and is operationally better for use (it is significantly faster in time) than solving Maxwell's equations.

The sources and the detectors used in experiments operate at frequency of 2.4 GHz, 
and detectors can measure the radiation intensity up to a maximum of about 200 measurements per seconds. However, we have verified that for our purposes such a large number of measurements is not necessary, it has been sufficient to have roughly an order of magnitude smaller measurement frequency.

During the first set of measurements (M1, see Fig. \ref{MjerniProstor}(a)), we have recorded the signal for 20 minutes on all three detectors, and the measurement conditions did not change. That is, first we recorded the noise - no one was in the room or near the room. After that, there was one person in the room and the signal was recorded for the next 20 minutes; the person moved randomly (not too fast but not uniformly), the person could also stop or make various movements. After that, we have recorded more measurements for 2, 3, 4 and 5 people inside the room. In this way, we have obtained 6 time series of 20 minutes each (approx. 200 measurements per second) for each detector.
We have also repeated the measurements the following day.
For the second set of measurements (M2), the configuration of the system was changed (see Fig. \ref{MjerniProstor}(b)), and the measurements were taken in an analogous way as described for the M1.

\subsection{Experiments: counting the number of persons}

In the second part of the research, we have aimed at finding an optimal approach to the more complicated problem of counting people based on the measurements of Wi-Fi signal strength.
It is important to emphasize that for this analysis we have conducted experiments wither larger number of detectors than in the first part, in order to generate more accurate predictive models.

\begin{figure}[!t]
\centerline{\includegraphics[width=0.45 \columnwidth]{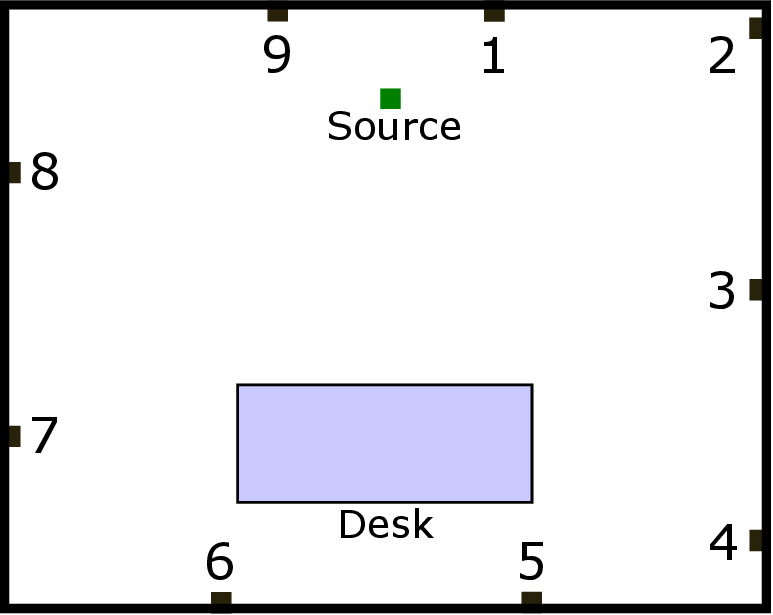}}
\caption{Floor plan of the experimental setup for Wi-Fi signal measurements with one source and 9 detectors. The source (and detectors) were placed at a height of 0.5 (and 1) m above the ground. The dimension of the measurement room is $4 \times 4.5$ m$^2$.}
\label{GeometrijaProstor}
\end{figure}

The geometry of the experimental setup (with radiation source and detectors) is presented in Fig. \ref{GeometrijaProstor}.
We had a total of nine detectors placed in different places within the room, which recorded Wi-Fi signal strength (RSSI) over time.
We have made measurements in the presence of one, three, five, seven or nine people. People were either at rest or in motion. For each cycle with a defined number of people, we collected data for twenty minutes: ten minutes with people standing still and ten minutes when people are moving. We also collected the noise data, without people present, in order to have reference data on the signal strength in the room.

We first divided the time series obtained by measurements into intervals of 20 seconds. In this way, for each measurement that lasted 20 minutes, we obtained a total of 60 different subintervals. Considering that we had a total of 5 such measurements (for the number of persons 1, 3, 5, 7 or 9), we finally had a set of 300 different time intervals at our disposal.

\section{Results and discussion}

\subsection{Detecting the presence of persons}

In real systems, it is often the case that we have an imbalance in the amount of data, i.e. much more data of only one type (e.g. noise) compared to others. In that case, we can consider the majority of the data as normal system behavior, and the other data as anomalies. Since we have few anomalies, and they can generally be of different shapes, a useful approach is to look for algorithms that will be trained exclusively on data that describe the normal state of the system (in our case, noise), i.e., without the presence of anomalies.
The key task is to determine whether someone was in the room during the tested time interval. The classification label for the detection of presence is a binary variable: 0 – noise (no people), or 1 – there is someone in the room.

It has already been clear to us, from several small preliminary signal samples, that signal fluctuations depend in some way on the presence of people in the room.
For this reason, we have studied the moments of the distribution of all signals, that is, we have calculated the statistical indicators of the signal: mean value, standard deviation, skewness, kurtosis; and also the Fourier spectrum.
When doing such analysis, it was necessary to decide how long each studied time series should be, given that the Fourier spectrum and other parameters oscillate somewhat in time even when the signal recording conditions do not change. We have been guided by the idea that within 10 or 20 seconds one would like to know the number of people in the room, or whether there is someone in the measurement area or not. However, it is possible to imagine applications where it is sufficient to know how many people have stayed in a certain space for a long time interval on average.

Let us imagine that from one measured time series of length $T=20$ min we want to extract quantity $A$ (eg., standard deviation, etc.) by the following procedure. Each time series is split into shorter time series of $\tau$ seconds. Each method, depending on the goal of the application, has a corresponding optimal value of $\tau$ (in what follows, we set $\tau = 20$ s). In this way, from one series we get $N_{\tau}=\text{floor}(T/\tau)$ shorter time series ($(T/\tau)$ is rounded to the nearest whole number towards zero). Then, for each series of $\tau$ seconds, we calculate the value of $A$; we can do this for each detector separately. (However, we note that it is also possible to first adequately combine the data from different detectors (eg. by multiplication) and then calculate the magnitude of $A$ for the combined series.)

\subsubsection{Information extraction from statistical parameters}

\begin{figure}[!t]
\centerline{\includegraphics{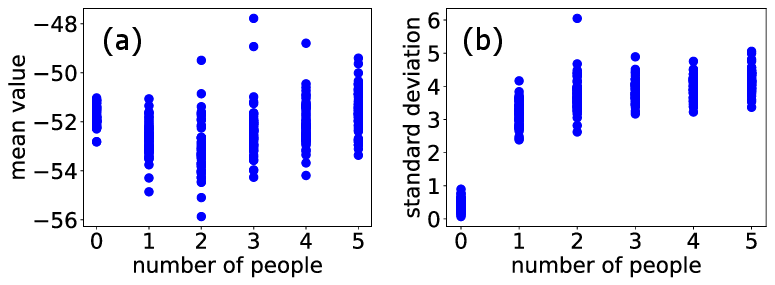}}
\caption{Mean value (a) and standard deviation (b) of the RSSI signal as a function of the number of people in the room covered by the Wi-Fi signal. Noise is marked with the number 0. Total length of the measurement is 20 minutes for each number of people, and the signal is split into series of 20-second samples. The mean value and standard deviation were calculated and plotted for the 20-second signal samples. There is no correlation between the mean value and the presence of people; however, a correlation is observed between the deviation and the presence of people.}
\label{SlMeanSTD}
\end{figure}

To begin with, we analyze moments of distribution for the measured signals in order to develop appropriate algorithms for given applications.
For the first set of measurements (M1), each time series of $20$ minutes is split into shorter time series of $\tau=20$ seconds.
In Fig. \ref{SlMeanSTD} we show dependence of the mean value of the signal and the standard deviation on the number of people.
It is clear from the figure that the standard deviation is correlated with the presence of people and can be used as a discriminating factor in decision-making. On the other hand, the mean value seems not to be correlated and its relevance is questionable.

We interpret this observation as follows: if people are in the room and are moving, the radiation is reflected from them and causes larger signal fluctuations which are seen in the standard deviation of the signal. 
That is, we find that the values of standard deviation for the noise measurements (0 people) are significantly smaller than its values when people were present in the room. Therefore, we conclude that (for this configuration) the standard deviation could be used to detect the presence of people. The advantage of this method is that it is not necessary to calibrate (train) the predictive model with measurements when people are present, but only with noise measurements. We have validated this conclusion for all three detectors used.
In addition, we note that the standard deviation cannot be used (as the only parameter) to extract precise information about the number of people in the room. (We will later discuss this problem with machine learning approaches.)

On the other hand, the mean value is affected by the position of the person in such a way that the signal value increases through reflection and decreases through absorption, so that there is no one-way influence of the presence of people in the room on the mean value of the signal.
We have also calculated other statistical quantities, i.e., the skewness and the kurtosis (not shown). We had noted that noise values for these parameters are more scattered when compared to the signal with people present, however, there is no clear discriminating procedure by which these parameters would be used for the binary decision-making application (i.e., to decide on presence or absence of people). Therefore, we do not focus on these parameters in the present discussion.

\paragraph{Predictive model (algorithm) for binary classification based on standard deviation of the noise - Method 1}
The input data for the algorithm are only the noise measurements. The (average value of) standard deviation $\langle \sigma \rangle$ has been extracted for all three detectors (we focus first on M1 configuration):
$\langle \sigma \rangle_{D1} = 0.4317$, 
$\langle \sigma \rangle_{D2} = 0.4214$, and 
$\langle \sigma \rangle_{D3} = 0.4787$. 


From Fig. \ref{SlMeanSTD} it is clear that if the standard deviation (of a 20-seconds time series) is greater than a factor $f \cdot \langle \sigma \rangle$, then there is almost certainly someone in the room. Now it is necessary to determine the optimal value for the factor $f$. Based on our measurements, we have chosen the value $f=2.2$. If the detector D$j$ ($j=1,2,3$), within $\tau=20$ s, has a standard deviation greater than $2.2 \langle \sigma \rangle_{Dj}$, that detector indicates that someone is in the room. Otherwise, there is no one in the room. Since we have three detectors, it is necessary to make a final decision from their output results. We have given every detector the same weight in the decision. That is, if at least two detectors suggest that someone is in the room, the decision is that someone is in the room.

We tested the algorithm on measurements from the M1 set
(with 1 to 5 people; and with 0 to 4 people, taken the following day).
Prediction accuracy of the algorithm has been $100\%$ for all measurements, with a single exception of one measurement with one person present and recorded one day after the noise measurement used for training, for which we found $90\%$ accuracy.

\begin{figure}[!t]
\centerline{\includegraphics[width=0.5 \columnwidth]{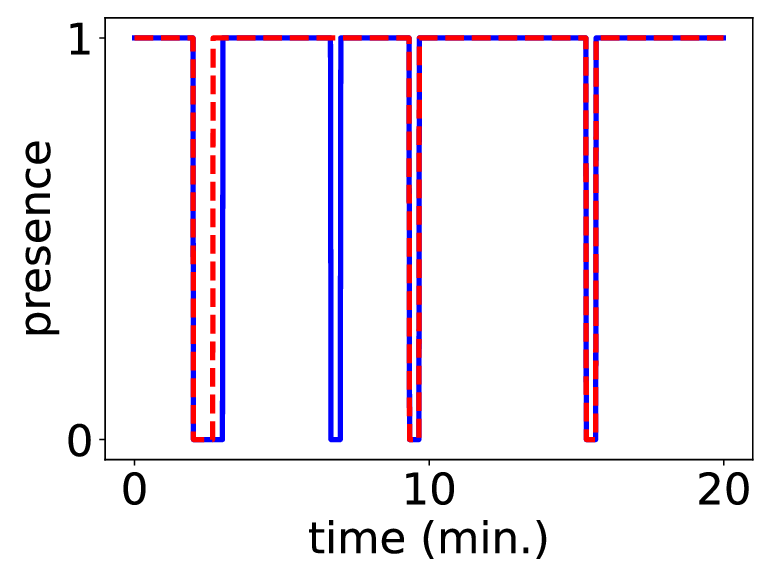}}
\caption{Prediction of two algorithms for binary classification (blue: Method 1, red: Isolation Forest) for the measurement made one day after the calibration and with one person present in the measurement room ($1$ stands for the presence being detected, and $0$ for the noise). This was the only measurement in the M1 set (configuration with both source and detectors inside the room) for which the algorithms could not reach $100\%$ accuracy (here: $90\%$ for Method 1, and $93\%$ for Isolation Forest).}
\label{SlM-121}
\end{figure}

Let us focus now only on that weaker result. Fig. \ref{SlM-121} shows the algorithm predictions for that measurement (with blue line for the method we presently discuss). There is only one longer time interval of about 1 minute when the method missed the correct value. We conclude that even for that worst measurement, the algorithm can predict within a longer time interval whether someone is in the room. Furthermore, it is necessary to take into account the long-term stability of the algorithm, i.e, examine whether the algorithm becomes worse if it is not recalibrated for days.
%


We have also trained and tested the algorithm on measurements from the M2 set.
From Table \ref{TabSer2} we can see that the results are somewhat worse than for M1, which we associate with a different position of the source (see Fig. \ref{MjerniProstor}). However, they were satisfactory for all measurements, except for the M-211 (one person present).
\begin{table}
\begin{tabular}{|l| c | c| c| c |c| c | c|} 
 \hline
Measurement  &  M-211	& M-212	&M-213	&M-214	&M-22X&	M-22Y	& M-22Z   \\
 \hline
Number of persons &  1 & 2 & 3	& 4 &1 & 2 & 3 \\ 
 \hline
 Method 1 & $45\%$	& $88\%$	&$95\%$	&$100\%$	&$83\%$	&$97\%$	&$100\%$\\ 
 \hline
 Method 2a & $72\%$	& $98\%$	&$100\%$	&$100\%$	&$100\%$	&$98\%$	&$100\%$\\
 \hline
 Method 2b & $67\%$	& $97\%$	&$100\%$	&$100\%$	&$100\%$	&$98\%$	&$98\%$\\
 \hline
 Isolation Forest & $72\%$	& $97\%$	&$100\%$	&$100\%$	&$100\%$	&$100\%$	&$100\%$\\
 \hline
\end{tabular}
\caption{\label{TabSer2}
Prediction accuracy of four different algorithms used for presence detection for the measurement set M2; detection is more challenging since the radiation source is placed outside of the measurement room.}
\end{table}


Our conclusion on the usability of the algorithm is as follows. This algorithm can be used for practical purposes for precise detection of whether or not there are people in the Wi-Fi signal field. The precision of the algorithm can be extremely high, but it depends on the position of the source and the detector. Importantly, in longer time intervals (for example, averaged results for 1-2 minutes), the algorithm practically makes no errors.


The notion that standard deviation is the statistical parameter which is well correlated with the presence (or absence) of people can be implemented in several other ways. In what follows we provide two specific approaches.

\paragraph{Predictive model (algorithm) for binary classification - Method 2a}

We have already seen that the noise deviation is significantly smaller than the signal deviation when people are present in a measurement area. By multiplying the deviations at different detectors, such differences can become even larger (more pronounced).
For each detector, we determine standard deviation of the total signal (for the whole measurement of 20 minutes) and then define the product of these deviations as the correlated noise deviation (for three detectors, we multiply three deviations). After that, we analyze the test signal (of length $\tau=20$ s): we find the standard deviation for each detector, and their product defines the correlated deviation.
A simple protocol for decision-making can be defined as follows: if the correlated deviation of the test data is by a factor 3 (or more) greater than the correlated noise deviation, then the algorithm predicts that (one or more) persons are present in the room. Otherwise, there are no people in the room.


\paragraph{Predictive model (algorithm) for binary classification - Method 2b}

We first multiply noise measurements from all three detectors at the same time (during the entire period, in our case 20 minutes) and find the standard deviation of such data set ($\sigma_{0,D1 D2 D3}$). For testing data (of length $\tau=20$ s) we calculate the correlated data by multiplying the measurements from all 3 detectors at the same time.
Protocol for decision-making is defined as follows: if the standard deviation of the tested correlated data is by a factor 3 (or more) greater than $\sigma_{0,D1 D2 D3}$, the algorithm says that (one or more) persons are present in the room. Otherwise, there are no people in the room.

\subsubsection{Information extraction using machine learning}


Finally, we have conducted the analysis of our measurements with machine learning techniques.
We have extracted features which characterize individual time series using the readily available software packages \cite{tsfresh}, and then treated them as input variables for the classification problem.
We have employed several well-known algorithms which can be used for semi-supervised learning (when only the noise data is used for algorithm training), and have chosen the Isolation Forest algorithm to be the well suitable for our data sets (other possible choices include One-Class Support Vector Machine, Local Outlier Factor, Elliptic Envelope, etc.)
This approach can also serve as a benchmark for the quality of previously introduced statistical methods.


%

\subsubsection{Comparison of different methods}

For the source and detector configuration of M1 data sets (when both source and detectors are inside the measurement room), all algorithms give approximately equal results, and they are extremely precise.
Specifically, prediction scores have not been perfect (but still of the order of $90\%$ or more) in a single case: when one person was inside the room and during the measurement taken one day after the noise measurements used for calibration (algorithm training), see Fig. \ref{SlM-121}. We conclude that the sensitivity (i.e., precision) can change over time, so it is necessary to calibrate the measuring system (parameters used by the algorithms) every day (calibration can be preferably done at night when there is no one in the room).

For the M2 configuration (when the radiation source is placed outside the room), the prediction results of all algorithms are generally somewhat worse than for the M1, and they are summarized in Table \ref{TabSer2}.
(Numbers presented may marginally change depending on data sampling rate, the choice of time series length $\tau$, or machine learning (for the Isolation Forest) parameters.)
Notably, algorithms are of limited use in the M2 configuration when only a single person is present in the measurement room (in particular, Method 1 gives worse results on measurements). 

Obviously, we find that the precision of the measurements (and the prediction accuracy of algorithms) depends on the configuration of the source and the detectors. (Throughout this discussion we have relied on data from three detectors, but in real-life situations even a single detector may be sufficient.)
Finally, we conclude that with the daily noise calibration, it is possible to employ practical algorithms that will precisely detect people presence in the measurement field, by using only the RSSI signal. Algorithms have correlated errors, so unfortunately they cannot be used as vectors in the sense that when one fails we use the other one.

\subsection{Counting the number of persons}

As already established, determining the number of people in the observed room based on the strength of the electromagnetic (Wi-Fi) signal can be defined as a classification problem in the field of machine learning (here, the classification label is the number of people in the room, and we will discuss supervised learning algorithms). For each of the 300 time series, we have performed the extraction of relevant featured that characterize it (as noted, we did this using a software package \cite{tsfresh}); these features then represent input for specific algorithms. In accordance with the standard approach, collected data can be divided into two sets, for training and for testing of the algorithm. As a result, we get predictions of classification labels, i.e., the predicted number of people in the measurement room during the tested time interval.

We have made the division into a training set and a testing set by using cross-validation: this means that the data is divided into $k$ subsets, where $(k-1)$ subsets are used for training and the remaining one for testing. Our set consists of a total of 300 elements, and for this analysis we have chosen the value of the cross-validation parameter $k=3$, which meant that the tested samples represent one third of the available data. In this way, training of each algorithm (and then its testing) can be carried out three times (for convenience, we label them as Run 1, Run 2 and Run 3).

The Table \ref{AnalysisExperimentMultipleDetectorsTable} shows the prediction success of three well established classification algorithms, based on measurements data from all 9 detectors.
\begin{table}
\begin{tabular}{|l |c | c| c|c|} 
 \hline
Algorithm  &   KNeighbors		&Decision Tree	&Random Forest	 \\
 \hline
Run 1 &  $95\%$ & $92\%$	& $99\%$  \\ 
 \hline
Run 2 &  $96\%$ & $91\%$	& $98\%$  \\ 
 \hline
Run 3 &  $93\%$ & $97\%$	& $100\%$  \\ 
 \hline
Average & $95\%$ & $93\%$	& $99\%$  \\ 
 \hline
\end{tabular}
\caption{\label{AnalysisExperimentMultipleDetectorsTable}
Prediction accuracy of three different machine learning algorithms (K-nearest neighbors, Decision Tree, and Random Forest) for classification problem of determining the number of people in the room, and for the test data recorded on nine detectors. Algorithms have been employed with cross-validation parameter $k=3$ (2/3 of data for training, 1/3 for testing), resulting in three separate runs (models).}
\end{table}
We note that all three selected algorithms are excellent in predicting the number of people in the room: K-Nearest Neighbors (KNeighbors), Decision Tree, and Random Forest (it also seems that this one should be highlighted in terms of success). We have also made preliminary tests with neural network algorithms (not shown here); however, this approach could be more useful (and successful) with larger data sets which we do not presently have (and further optimized by changing the parameters of the neural network).
 
\begin{figure}[!t]
\centerline{\includegraphics[width=0.65 \columnwidth]{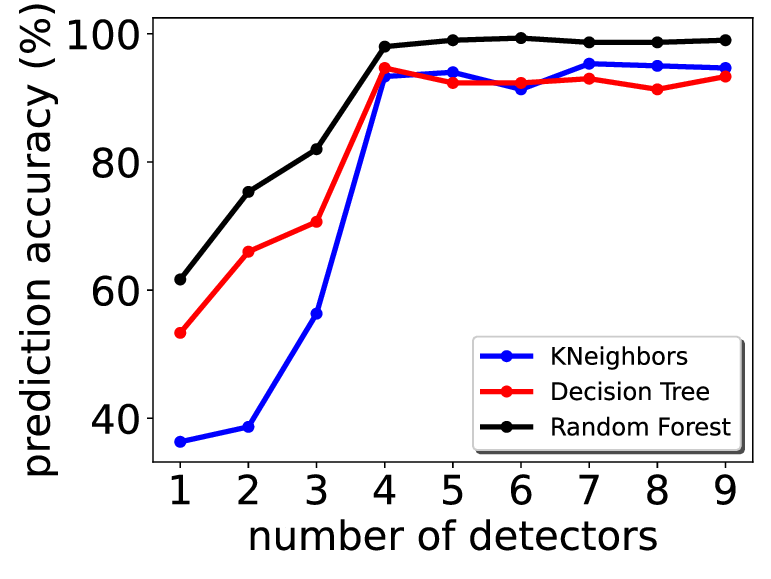}}
\caption{Prediction accuracy of three different algorithms (K-nearest neighbors, Decision Tree, and Random Forest) for the classification problem of counting the number of people inside the measurement room, plotted as a function of the number of detectors used for analysis.}
\label{AnalysisExperimentMultipleDetectors}
\end{figure}

Given that the prediction results based on the data of all 9 detectors are excellent, below we will demonstrate how these results depend on the number of detectors. The goal of this check is to determine the minimum number of detectors in the room required to obtain satisfactory prediction accuracy.
Fig. \ref{AnalysisExperimentMultipleDetectors} shows prediction accuracy of the chosen algorithms as a function of the number of detectors used in the analysis. It can be seen that the prediction success initially increases with the increase in the number of detectors, but that already at the number of 4 detectors, successful algorithms reach excellent prediction power, which changes only marginally with an additional increase in the number of detectors. Most notably, the Random Forest has performed with accuracy of $98\%$ when four detectors (labeled as 1 to 4 on Fig. \ref{GeometrijaProstor}) are used, and reaches $99\%$ already with five detectors (1 to 5 on Fig. \ref{GeometrijaProstor}). Obviously, these results may slightly change depending on the spatial configuration of the detectors used inside the measurement area. In particular, we do not discuss here generalizations to more complex indoor environements, such as multiple rooms of odds shapes and/or greater size, which might require more careful choice of detector positions, or even additional detectors to get the same level of prediction accuracy. Yet, even in such scenarios, our RSSI-based approach may provide a good alternative to more complex (and less practical) methods.

Finally, we emphasize here the most important results in this part of the research. Machine learning algorithms have shown an excellent ability to predict the number of people inside the measurement room based on the Wi-Fi (RSSI) signal, in an experimental setup with up to 9 people (and a total of 9 detectors). In particular, we have shown that the algorithms can achieve exceptional prediction success when analyzing experimental data even with only few detectors present in the room. For example, it is worth noting that the Random Forest algorithm performed with $98\%$ ($99\%$) success rate for the data collected on four (five) detectors for our classification problem. As an outlook, this study could be extended to an even larger experimental setup, with more people (and possibly more detectors); exact counting becomes more demanding, but finding an optimal (and cost-effective) approach will surely depend on specific applications.

\section{Conclusions}

In summary, we have measured the strength of the Wi-Fi signal (i.e., the RSSI value) on various detectors with the aim to determine the presence (or absence) and the number of people in an observed environment.
Importantly, persons involved do not actively cooperate (signal strength on their Wi-Fi devices is not recorded).
Our subsequent analysis has shown that the presence of one (or more) persons can be established with only a single quantity: the standard deviation of the RSSI value, and a small number of detectors (up to three in our experiment).
This understanding can be implemented in several ways, all of which lead to decision-making algorithms which can be of high precision when making a binary prediction (presence or absence).
They compare well with more advanced methods (such as machine learning algorithms), yet remain relatively simple.
However, we have also established that the same algorithms can be of limited use in experimental setup when the source of the Wi-Fi signal is located outside the observed environment.

As for the more challenging problem of counting the number of persons, more detectors are generally needed (we had a total of nine detectors placed in an indoor environment) and the data analysis has been performed by employing standard machine learning algorithms (a single feature, such as the standard deviation of the RSSI, does not perform well in this classification problem). For our experimental setup (with up to nine people), we have verified that algorithms were counting the number of people with high accuracy. For example, the Random Forest algorithm has had the average success rate of $99\%$ (the classification problem was designed to discriminate between labels 1, 3, 5, 7, and 9 for the number of people). Importantly, we have also shown that the excellent prediction accuracy of ML algorithms can be achieved when taking into account data from only few detectors. This final observation has also been motivated by possible real-life practical applications, where employing a minimal yet accurate system, is very beneficial. \\

%

\small{

\noindent {\bf{Abbreviations}} \\
\noindent
\begin{tabular}{@{}l l}
RSSI & Received Signal Strength Indicator \\
CSI & Channel State Information \\
PWR & Passive Wi-Fi Radar \\
LOS & Line of Sight \\
SVR &  Support Vector Regression \\
MIMO & Multiple Input - Multiple Output \\
OFDM & Orthogonal Frequency Division Multiplexing \\
ML & Machine Learning \\
KNeighbors & K-Nearest Neighbors \\
\end{tabular} \\

\noindent {\bf{Funding and Acknowledgments}} \\
\noindent 
The work was supported by the European Union through the European Regional Development Fund (Grant
No. KK.01.2.1.02.0016).
The authors would like to thank all colleagues, either at the University of Zagreb or at the Aduro Idea, who helped in organization of the experiments and with their presence when the experimental data were being taken. \\

\noindent {\bf{Author contributions}} \\
\noindent 
D.J., S.D. and H.B. performed theoretical and numerical analysis
of experimental results.
A.I., D.R., S.N., M.K., and F.J performed experiments.
N.R. and H.B. conceived the project and supervised the work.
D.J. wrote the manuscript.
All the authors discussed the results
and contributed to this work. \\

\noindent {\bf{Availability of data and materials}} \\
\noindent
The datasets used and/or analyzed during the current study are available from the corresponding author on reasonable
request.\\

}

\normalsize
\noindent{\bf{Declarations}} \\

\small{

\noindent {\bf{Competing interests}} \\
The authors declare that they have no competing interests. \\

}

\end{document}